\documentstyle[pre,aps,float,epsf,multicol]{revtex}
\begin{document}
\draft
\title{\bf Comment on Renormalization Group Study of the $A+B\to0$
Diffusion-Limited Reaction}
\author{Benjamin P. Lee}
\address{Polymers Division, National Institute of Standards and Technology,
Gaithersburg, MD 20899, USA}
\author{John Cardy}
\address{All Souls College and \\ Theoretical Physics, University of 
Oxford, 1 Keble Road, Oxford OX1 3NP, United Kingdom}
\date{December 6, 1996}
\maketitle
\begin{abstract}
A recent argument of Oerding shows that our calculation of the quantity
$\Delta$, which determines the amplitude of the asymptotic decay of the
particle density in $2<d<4$, was in error. Instead it is simply given by
$\Delta=n_0$, the initial density, for uncorrelated initial
conditions.
\end{abstract}
\pacs{Key Words: Diffusion-limited reaction; renormalization 
group; asymptotic densities}

\begin{multicols}{2}

In recent years much progress has been made in applying field-theoretic
renormalization group methods to models of diffusion-limited chemical
reactions.  One such example was a study by the present authors of
the $A+B\to\emptyset$ annihilation reaction with random initial conditions
and equal particle densities \cite{LC}.  Our purpose here is to
correct a mistake contained in this work, and to revise the conclusions
accordingly.  In particular, we now find for spatial dimensions $2<d<4$
that the $A$, $B$ particle densities decay in time with the universal form
\begin{equation}\label{den}
\langle a(t)\rangle=\langle b(t)\rangle\sim{\sqrt{n_0}\over\pi^{1/2}
(8\pi)^{d/4}}(Dt)^{-d/4}
\end{equation}
where $\langle a(0)\rangle=\langle b(0)\rangle=n_0$, and $D$ is the 
diffusion constant.  This amplitude is exactly that obtained in the
approximate calculation of Toussaint
and Wilczek \cite{TW}, and the initial density dependence is the same
as found by Bramson and Lebowitz in the limit of an instantaneous 
reaction \cite{BL}.  In our previous study we identified 
nonuniversal corrections to the amplitude in (\ref{den}) 
of order $n_0^{d/2}$  \cite{LC}.  However,
in analogy with an argument of Oerding \cite{Oerding} for the same
reaction in a random shear flow,  it can be shown quite generally that 
the sum of all such corrections in fact vanishes. We now briefly
paraphrase his arguments in the context of the present problem.

In general one begins with a master equation description of the 
diffusive and reactive particle dynamics.  This is
mapped to a field theory using by now standard techniques
\cite{LC,Doi,Peliti}.  It is convenient to replace the fields
$a(x,t)$ and $b(x,t)$ with their sum ($\phi$)
and difference ($\psi$), resulting in the action \cite{oops}
\begin{eqnarray}\label{action}
S=\int d^dx\>dt\biggl\{&&\bar\phi(\partial_t-\nabla^2)\phi+
\bar\psi(\partial_t-\nabla^2)\psi+\lambda_1\bar\phi(\phi^2-\psi^2)\nonumber\\
&&+\lambda_2(\bar\phi^2-\bar\psi^2)(\phi^2-\psi^2)-n_\phi\delta(t)\bar\phi
\biggr\}.
\end{eqnarray}
The relation between the couplings in (\ref{action}) and the original 
master equation parameters is given in ref.\ 1.  Note that
the diffusion constant is absorbed into
a rescaling of time, and the
conjugate fields $\bar\phi$ and $\bar\psi$ are introduced in the
mapping.

The renormalization group (RG) analysis reveals that the critical dimension
for the coupling $\lambda_2$ is $d_c=2$, i.e. $\lambda_2$ is irrelevant for
$d>2$.  Since the dynamic RG relates
evolution in time to renormalization group flows, one finds
the asymptotic behavior of the theory is given for $d>2$ by
an effective action of the form \cite{oops}
\begin{eqnarray}\label{eff_action}
S_{\rm eff}=\int &&d^dx\>dt\biggl\{\bar\phi(\partial_t-\nabla^2)\phi+
\bar\psi(\partial_t-\nabla^2)\psi\\
&&+\lambda_{\rm eff}\bar\phi(\phi^2-\psi^2)
-\delta(t)\Bigl[n_\phi\bar\phi+{\textstyle{1\over 2}}\Delta\bar\psi^2
+\ldots\Bigr]\biggr\}\nonumber.
\end{eqnarray}
Under the renormalization flows for $d>2$ 
one has $\lambda_1\to\lambda_{\rm eff}$,
and, more importantly, new initial terms are generated such as
${1\over 2}\Delta\bar\psi^2$.  While many such
terms may appear and be relevant for $d<4$, it was shown 
in ref.\ 1 that for $2<d<4$ 
the asymptotic behavior of the density and correlation functions is 
determined solely by $\Delta$.  Hence we now focus on the generation of
this coupling. (For $d>4$ one obtains
the rate equation result $\langle a(t)\rangle\propto 1/\lambda_{\rm eff}t$.)  

Denote the sum of all diagrams in an expansion of (\ref{action})
which terminate at time $t$ with two external $\psi$ lines
by $\bar\psi(t)^2\Pi(t)$. This is illustrated
schematically in fig.\ 4(a) of ref.\ 1.  Since $\Pi(t)$ is 
damped for times $t\gg 1/n_\phi\lambda_1$, the replacement
\begin{equation}
\bar\psi(t)^2\Pi(t)\simeq\bar\psi(0)^2\delta(t)\int_0^\infty dt'\Pi(t')
\equiv{\textstyle{1\over 2}}\Delta\bar\psi(0)^2\delta(t)
\end{equation}
is valid for calculating asymptotic quantities, and serves to define 
$\Delta$.  

The sum of all such diagrams containing no loops is given below in 
fig.\ 1(a), and yields a contribution $\Delta^{(a)}=n_0$, the initial
density.  Here, following the notation of ref.\ 1, the dashed line
represents the $\psi$ propagator, the heavy solid line is the $\phi$ 
classical (tree-level) response function, which is the $\phi$ propagator 
dressed 
by the initial density, and the wavy line is the classical density.
Three- and four-point vertices have coupling constants $\lambda_1$ and 
$\lambda_2$, respectively, with signs which can be determined from 
(\ref{action}).  In ref.\ 1
we assumed that all other diagrams in $\Pi(t)$, except those which
dressed the $\lambda_2$ vertex, could be accounted
for by taking $\lambda_1\to\lambda_{\rm eff}$ in the tree level
diagrams.  This left $\Delta^{(a)}$ unchanged, and the corrections
were found to yield an expansion in powers of $n_0$ of the form 
$\Delta=n_0-C'n_0^{d/2}+\ldots$, with $C'$ nonuniversal
(see equation (1.5) in ref.\ 1).
However, Oerding has shown that $\Delta$ can be calculated
exactly from the full action (\ref{action}), with the result \cite{Oerding}
\begin{equation}\label{Delta}
\Delta=n_0.
\end{equation}  
Since $\Delta$ itself is not renormalized \cite{LC}
this must be the correct value for the effective action (\ref{eff_action}),
in contradiction with the previous result.

\begin{figure}
\narrowtext
\epsfxsize=\hsize\epsfbox{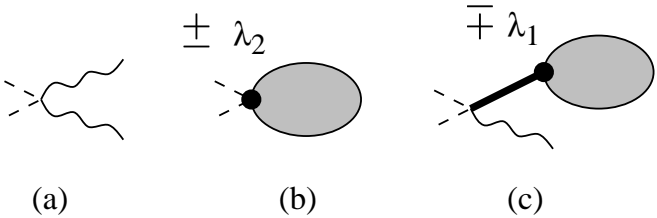}
\medskip
\caption{Diagrams which contribute to $\Pi(t)$, hence to $\Delta$.  The
symbols are defined in the text and in ref.\ 1.}
\end{figure}

Oerding's result follows from
observing that all diagrams in $\Pi(t)$ belong to one of the three groups
in fig.\ 1. 
The shaded area represents the same set of diagrams in fig.\ 1(b) as 
in fig.\ 1(c).  These may be attached to the 
highlighted vertex (represented by a dot) by
either two $\psi$ or two $\phi$ lines: in either case, the sign of
the $\lambda_2$ vertex in fig.\ 1(b) is the opposite of the $\lambda_1$
vertex in fig.\ 1(c), as can be confirmed by S (\ref{action}).  
Therefore, writing the contribution from fig.\ 1(b) as
\begin{equation}\label{Deltab}
\Delta^{(b)}=\lambda_2\int_0^\infty dt\>  f(t)
\end{equation}
implies that the contribution from fig.\ 1(c) is
\begin{eqnarray}\label{Deltac}
\Delta^{(c)}&=&\int_0^\infty dt {2\lambda_2n_\phi\over 1+n_\phi\lambda_1t}
\int_0^t dt'\left(1+n_\phi\lambda_1t'\over 1+n_\phi\lambda_1t\right)^2
(-\lambda_1)f(t')\nonumber\\
&=&-\lambda_2\int_0^\infty dt' f(t')
\end{eqnarray}
Hence $\Delta = \Delta^{(a)}$, which leads to  (\ref{Delta}) \cite{tinv}.

Using the correct and universal result for $\Delta$ in
the analysis of ref.\ 1 yields the asymptotic density (\ref{den}).
In ref.\ 1 we also derived the correlation functions and the density
in the case of unequal diffusion constants $D_A\neq D_B$.  Since these
results are expressed in terms of $\Delta$, they may be regarded as
correct and universal once the substitution (\ref{Delta}) is made.
However, we stress that these results are universal only for truly 
random initial conditions.  Correlations
initially present are expected to modify $\Delta$ in a non-universal
way.

We thank K.~Oerding for bringing this error to our attention.

\end{multicols}


\begin{references}
\bibitem{tinv}We note, however, that if $f(t)\sim 1/t$ for large $t$, the 
divergent parts of (\ref{Deltab}) and (\ref{Deltac}) cancel, and leave a 
finite correction to (\ref{Delta}).
\end{references}
\end{document}